\documentclass[]{fairmeta}

\usepackage{amsmath,amsfonts,bm}

\def\eqref#1{equation~\ref{#1}}

\def\1{\bm{1}}

\DeclareMathAlphabet{\mathsfit}{\encodingdefault}{\sfdefault}{m}{sl}
\SetMathAlphabet{\mathsfit}{bold}{\encodingdefault}{\sfdefault}{bx}{n}
\usepackage{url}
\usepackage{microtype}
\usepackage{amsmath, amsthm, amsfonts, amssymb,bbm}
\usepackage{dsfont} 
\usepackage{enumitem}
\usepackage{wrapfig}
\usepackage{pifont} 
\usepackage{array}

\usepackage{tabularx}
\usepackage{tablefootnote}
\usepackage{colortbl} 
\usepackage{makecell}
\usepackage{float}
\usepackage{threeparttable}

\usepackage{xspace}
\usepackage{xparse}
\usepackage[scientific-notation=true]{siunitx} 
\usepackage[ruled,vlined]{algorithm2e}
\usepackage{algpseudocode}
\usepackage{graphicx}
\usepackage{multirow} 
\usepackage{subcaption}
\usepackage{dsfont}  
\usepackage{annotate-equations}
\usepackage{lmodern}
\definecolor{Gray}{rgb}{0.5, 0.5, 0.5}
\definecolor{LightGray}{rgb}{0.8, 0.8, 0.8}
\definecolor{DarkGray}{rgb}{0.3, 0.3, 0.3}
\definecolor{SlateGray}{rgb}{0.44, 0.5, 0.56}
\definecolor{DimGray}{rgb}{0.41, 0.41, 0.41}
\definecolor{DarkSlateGray}{rgb}{0.18, 0.31, 0.31}
\newcommand{\graydelta}[1]{\textcolor{gray}{\footnotesize (#1)}}

\usepackage{hyperref}
\usepackage{url}
\usepackage{tcolorbox}          
\usepackage[font=small, skip=8pt]{caption}
\usepackage{subcaption} 
\usepackage{wrapfig}
\usepackage{xcolor} 
\usepackage{booktabs}
\usepackage{multirow}
\usepackage{enumitem}

\usepackage{float}

\usepackage{ifthen}  
\newboolean{set_me_to_remove_todos}
\setboolean{set_me_to_remove_todos}{false} 
\ifthenelse{\boolean{set_me_to_remove_todos}}{
    \newcommand{\pierre}[1]{}
    \newcommand{\tom}[1]{}
    \newcommand{\todo}[1]{}
}{
    \newcommand{\tom}[1]{{\color{purple} [\textbf{Tom}: #1]}}
    \newcommand{\pierre}[1]{{\color{blue} [\textbf{Pierre}: #1]}}
    \newcommand{\todo}[1]{{\color{red} [\textbf{TODO}: #1]}}
}

\definecolor{metablue}{HTML}{0064E0}
\definecolor{metafg}{HTML}{1C2B33}
\definecolor{metabg}{HTML}{F1F4F7}

\def\1{\mathbf{1}}

\def\V{\mathcal{V}}
\def\H{\mathcal{H}}
\def\Prob{\mathds{P}}

\newcommand{\eg}{\emph{e.g.},\@ }
\newcommand{\ie}{\emph{i.e.},\@ }

\newcommand{\logit}{\boldsymbol{\ell}}
\newcommand{\logpval}{\log_{10}(p)}
\newcommand{\pval}{p\textrm{-value}}

\definecolor{Gray}{gray}{0.95}

\newlength\savewidth

\newtheorem{proposition}{Proposition}

\newcommand{\sk}{\mathsf{s}}

\title{Detecting Benchmark Contamination \\ Through Watermarking}

\author[1,2]{Tom Sander}
\author[1]{Pierre Fernandez}
\author[1]{Saeed Mahloujifar}
\author[2]{Alain Durmus}
\author[1]{Chuan Guo}

\affiliation[1]{Meta FAIR}
\affiliation[2]{\'Ecole polytechnique CMAP}

\abstract{
Benchmark contamination poses a significant challenge to the reliability of Large Language Models (LLMs) evaluations, as it is difficult to assert whether a model has been trained on a test set. 
We introduce a solution to this problem by watermarking benchmarks before their release.
The embedding involves reformulating the original questions with a watermarked LLM, in a way that does not alter the benchmark utility.
During evaluation, we can detect ``radioactivity'', \ie traces that the text watermarks leave in the model during training, using a theoretically grounded statistical test.
We test our method by pre-training 1B models from scratch on 10B tokens with controlled benchmark contamination, and validate its effectiveness in detecting contamination on ARC-Easy, ARC-Challenge, and MMLU.
Results show similar benchmark utility post-watermarking and successful contamination detection when models are contaminated enough to enhance performance, \eg $p$-val $=10^{-3}$ for +5$\%$ on ARC-Easy.
}

\correspondence{\email{tomsander@meta.com}}

\begin{document}

\clearpage
\maketitle


\begin{figure*}[h]
    \centering
    \includegraphics[width=1.0\textwidth, clip, trim=0 3.3cm 1.2cm 0]{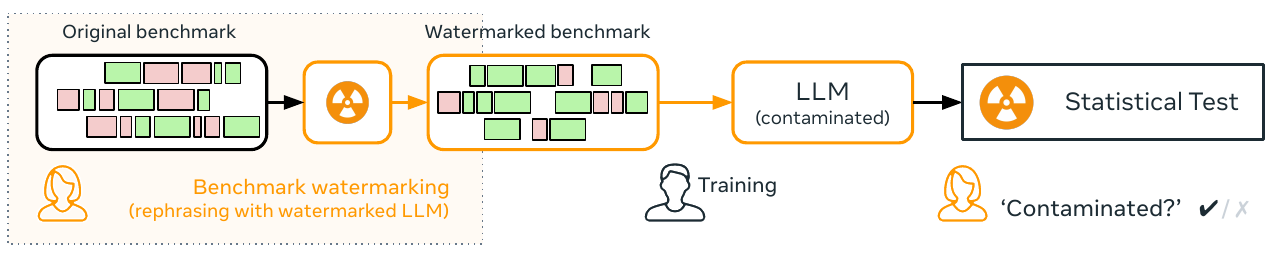}
    \captionsetup{font=small}
    \caption{
    Problem overview. 
    \emph{Alice} is a benchmark provider and wants to make sure that contamination on her benchmark can be detected with high confidence.
    Before release, she rephrases the original benchmark dataset while embedding a non-intrusive LLM watermark.
    This rephrasing does not change the utility of the benchmark.
    \emph{Bob} decides to train a model.
    The benchmark may contaminate Bob's model during training, either intentionally or unintentionally.
    Alice can give statistical evidence if her benchmark was used in training.
    }
    \vspace{-0.2cm}
    \label{fig:fig1}
\end{figure*}

\section{Introduction}\label{sec:intro}

In recent years, Large Language Models (LLMs) have demonstrated remarkable advancements in their capabilities~\citep{brown2020language, touvron2023llama}. 
This advancement places increasingly greater emphasis on proper evaluation to both inform the state of LLM research and to guide future developments.
To this end, a multitude of benchmark datasets such as (MMLU)~\citep{hendrycks2020measuring}, School Math 8K (GSM8K)~\citep{cobbe2021training}, and the AI2 Reasoning Challenge (ARC)~\citep{clark2018think}, or more recently GPQA~\citep{rein2023gpqa} and FrontierMath~\citep{glazer2024frontiermath},  are developed to measure the model's capabilities in terms of general or specific knowledge,  understanding, and scientific reasoning.

However, a significant issue that arises with these benchmarks is contamination. 
This problem can occur either intentionally, by training models directly on the benchmark datasets or their reformulated versions, or unintentionally, as these datasets become mixed with the vast amounts of data used during pre-training. 
For example, ~\citet{zhang2024careful} created a version of GSM8K with new questions similar in difficulty and form, and observed that many models show a significant drop in performance on them compared to the test set of GSM8k.
This challenges the reliability and validity of benchmark evaluations, as it becomes difficult to discern whether a model's performance is due to genuine improvement in capabilities or mere memorization.
Furthermore, determining whether a model has been trained on a specific benchmark is very challenging, as it boils down to the issue of dataset/membership inference which has been shown to be ineffective for LLMs in realistic scenarios~\citep{duan2024membership}: it necessitates a prior on how the model would behave if not contaminated, such as through a held-out set from the same distribution.
In the context of benchmarks, having a held-out set would render the problem moot, as simply evaluating on it resolves the issue.

To tackle this problem, we propose a novel strategy of embedding non-intrusive watermarks in the benchmark dataset before release.
Our approach is inspired by \citet{sander2024watermarking}, who demonstrated that slightly distilling a watermarked LLM can be reliably detected, as the model retains identifiable traces of the watermark.
We extend this idea to benchmark watermarking, and when possibly different tokenizers are used by the watermarking and the suspect models.
Our approach enables reporting both model performance on the benchmark and a reliable $p$-value as a contamination score, which relates to the False Positive Rate of the contamination test (see Proposition~\ref{proposition:fpr_p_val}).
If the reported $p$-value is low, the LLM's training data is likely contaminated with the benchmark dataset and the performance numbers should not be trusted as genuine.
Our method requires only access to an LLM capable of rephrasing benchmark questions; see \autoref{fig:fig1} for an overview.
Our main contributions are:

\begin{itemize}[leftmargin=*]
    \item \textbf{Rephrasing benchmark datasets with watermarking:} We use Llama-3-8B-Instruct to rephrase questions from MMLU, ARC-Easy, and ARC-Challenge benchmarks. By applying the red/green list watermarking technique from \citet{kirchenbauer2023watermark}, we show that rephrasing effectively incorporates watermarks while preserving benchmark integrity (\autoref{subsec:rephrasing}).
    \item \textbf{Extending watermark radioactivity:} We extend watermark radioactivity to a pre-training setup. 
    We train models with up to 1B parameters on 10B tokens, varying benchmark contamination levels, each benchmark with a different secret watermarking key $\sk$.
    For instance, our results show detection of contamination with a $p$-value below $10^{-5}$ when the accuracy is only inflated by $5\%$ on 5000 MMLU questions, indicating a one in 100k chance of error to falsely flag a model as contaminated, while correctly yielding $p$-values near $0.5$ for uncontaminated models (\autoref{fig:results_overview_arc_easy_detection} and \autoref{tab:contamination}).
    \item \textbf{New detection algorithm when different tokenizers are used.} \citet{sander2024watermarking} assumed that the same tokenizer was used for both the watermarking model and the suspect model. 
    We propose a new algorithm when a different tokenizer is used by the suspect model, and show that it comes with reasonable loss in contamination detection confidence (Algorithm~\ref{alg:reading_mode_different_tokenizers} in sec.~\ref{subsec:different_tokenizers} and sec.~\ref{subsec:results_tokenizers})
\end{itemize}

\vspace{-0.2cm}
\section{Related Work}\label{sec:related}


\subsection{Benchmark Contamination Detection}
Benchmark contamination is a significant concern in evaluating LLMs, as it can lead to unreliable assessments and unfair comparisons~\citep{singh2024evaluation, balloccu2024leak}. 
Although efforts are made to decontaminate pre-training corpora~\citep{brown2020language}, these methods are not foolproof~\citep{singh2024evaluation}.
The impact of contamination can be assessed by comparing training runs that differ only in the inclusion of contaminated batches:~\citet{jiang2024investigating} use this to show that even small models can exhibit improved benchmark performance due to small contamination.
Post-hoc analyses on the other hand identify score inflation by comparing performance on original versus similar questions~\citep{brown2020language, chowdhery2023palm}, but~\citet{yang2023rethinking} have shown that training on reformulated questions is enough to boost the performance on the original benchmark, so the difference in performance does not necessarily provide good correlational insights.

\citet{zhang2024careful} craft new questions from the same distribution as GSM8K and observed that most models show a significant performance drop on these compared to the GSM8K test set. 
This result highlights the contamination issue, but does not introduce a scalable solution to the problem.
In parallel, studying memorization in LLMs, such as regurgitating pre-training data, is also closely related~\citep{carlini2022quantifying, hartmann2023sok}.
Techniques like membership inference attacks~\citep{mireshghallah2022quantifying} and context-based completion checks~\citep{golchin2023data} attempt to approximate contamination without direct access to pre-training data, but their effectiveness is debated~\citep{duan2024membership}. 
Moreover, membership inference necessitates a held-out set from the same distribution, which, in the contexts of benchmarks, renders the problem moot.


\vspace{-0.2cm}
\subsection{Decoding-based watermarking \& Radioactivity}\label{subsec:related_kirch}

\paragraph{\textbf{Overview.}} 
Recent advancements in watermarking techniques for large language models (LLMs) involve altering either the probability distribution~\citep{kirchenbauer2023watermark} or the method used for sampling the subsequent token~\citep{aaronson2023watermarking, kuditipudi2023robust}.
Detection of these watermarks is influenced by the entropy of the generated text~\citep{christ2023undetectable, huang2023optimal}, so further investigations propose watermarking only sections with high entropy, especially in code~\citep{lee2023wrote}, while other studies explore ``semantic'' watermarks that rely on the semantic representation of the entire preceding text~\citep{liu2023semantic, liu2024adaptive, fu2024watermarking}.

\paragraph{\textbf{Green-list/Red-list watermark.}} Our work focuses on the watermarking scheme proposed by~\citet{kirchenbauer2023reliability}, which modifies the logit vector during token generation based on a context window of $k$ previous tokens and a private key $\sk$. 
Both are hashed to serve as the seed for a random number generator (RNG) to create a ``greenlist'' of $\gamma |\V|$ tokens, where $\V$ is the vocabulary of the tokenizer.
Logits of green tokens are incremented by $\delta$ to increase their sampling probability.
Detection involves repeating the greenlist computation for each token of a text, incrementing a score by 1 if the token is in the greenlist, and performing a statistical test on the cumulative score.
Under the null hypothesis $\H_0$  ``the text is not watermarked with that scheme'', this score follows a binomial distribution~\citep{fernandez2023three}.
A simple binomial test thus provides a $p$-value.


\paragraph{\textbf{Radioactivity of LLM watermarks.}}
\citet{sander2024watermarking} show that fine-tuning language models on LLM-generated watermarked question-answer pairs can be detected with high confidence, as the model retains traces of the watermark bias. 
The authors adapt the original watermark detection tests to detect watermark ``radioactivity''---a term first coined in ~\citet{sablayrolles2020radioactive} for image data---depending on the access to the suspect model and its training data.
In our work, we assume open-weight access to the LLM that is being evaluated, either because it is open-source, or because the contamination test is ran by the LLM owners themselves.
Moreover, the benchmark as well as the benchmark-specific watermarking key $\sk$ are both known during radioactivity detection.
In this case,~\citet{sander2024watermarking} suggest using what they call ``reading-mode''. 
This involves scoring all next-token predictions by forwarding the watermarked text in the suspect model.
This is detailed in our context in~\autoref{subsec:detection}.
Similar observations had been made in other scenarios.
For instance, \citet{gu2023learnability} demonstrate that LLM watermarks can be intentionally distilled. 
Additionally, \citet{zhao2023protecting} introduce a signal in generated text that can be learned by other LLMs trained on it, and \citet{jovanovic2024ward} investigate the concept of watermark radioactivity for RAG.

\vspace{-0.3cm}
\section{Method}\label{sec:method}

We first focus in section~\ref{subsec:rephrasing} on the task of rephrasing the questions of a benchmark dataset while embedding a watermark using the method proposed by~\citet{kirchenbauer2023reliability}.
Then, in section~\ref{subsec:detection}, we show how to detect if a language model was trained on the watermarked benchmark.
Finally, section~\ref{subsec:different_tokenizers} extends the  test if a different tokenizer is used by the suspect model.

\vspace{-0.1cm}
\subsection{Inserting watermark through question rephrasing}\label{subsec:rephrasing}

We use an instruct language model, denoted as $LM_{\text{rephrase}}$, which is assumed to be capable of rephrasing each question in the benchmark test set such that the rephrased version is logically equivalent to the original.
This is a pretty light assumption as the task of rephrasing is considerably easier than answering the question~\citep{deng2023rephrase}.
$LM_{\text{rephrase}}$ generates token per token and at each step, takes as input a context, which is the concatenation of the system prompt, rephrasing instruction, the question to rephrase and the reformulation generated so far.
Everything is tokenized into a sequence $\left( x^{(1)}, \ldots, x^{(t-1)} \right) \in \V^{t-1}$, where $\V$ is the vocabulary of the tokenizer.

$LM_{\text{rephrase}}$ outputs a logits vector $\logit^{(t)} \in \mathbb{R}^{|\V|}$.
The watermark embedding modifies $\logit^{(t)}$ based on a secret key $\sk$ (one per benchmark) and the watermark window 
$\left( x^{(t-k)}, \ldots, x^{(t-1)} \right) \in \V^k$

\begin{figure}[b!]
    \vspace{-0.2cm}
    \centering
    \begin{minipage}{0.48\textwidth} 
        \begin{tcolorbox}[
            colframe=metablue, 
            colback=white, 
            width=\textwidth,
            left=2mm, right=2mm, top=2mm, bottom=2mm
        ] 
            {\footnotesize
            \textbf{System prompt + instruction:} \\``You are a problem rephrasing assistant [...]''
            \\[4pt]
            \textbf{Question:} ``The rate of acceleration of an object is determined by the mass of the object and''
            \\[4pt]
            \textbf{Rephrased with watermark ($\delta=4$):}\\
            ``What factor, aside from an object's mass, determines its acceleration?'' ($73\%$ of green tokens)
            }
        \end{tcolorbox}
        \subcaption{Embedding - benchmark rephrasing}
        \label{fig:example_answers_main}
    \end{minipage}
    \hspace{0.02\textwidth} 
    \begin{minipage}{0.48\textwidth} 
        \centering
        \includegraphics[width=\textwidth]{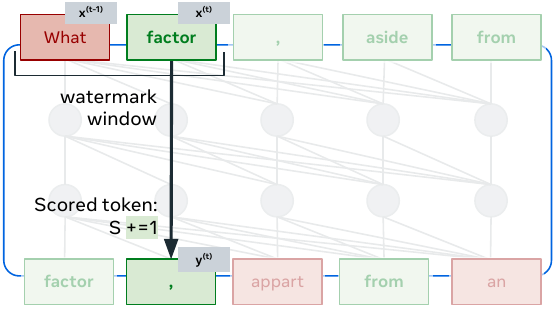} 
        \subcaption{Detection - statistical test}
        \label{fig:method_overview}
    \end{minipage}
    \caption{Method description. (Left) Watermarking the benchmark's questions using an LLM, as detailed in~\autoref{subsec:rephrasing}, with an example from ARC-easy. 
    The quality of the question is maintained despite strong watermarking. (Right) Reading mode, as detailed in ~\autoref{subsec:detection}.
    The upper sequence is the watermarked question, and the tokens bellow are the top-1 predictions from the suspect model ($y^{(t)}$ in~\autoref{eq:watermark_score}).}
    \vspace{-0.3cm}
    \label{fig:method_main}
\end{figure}

Specifically, following the method of \citet{kirchenbauer2023reliability} detailed in~\ref{subsec:related_kirch}, a secret-key cryptographic function hashes $\sk$ as well as the the watermark window, which serves as a seed for a random number generator used to create a pseudo-random ``greenlist'' of tokens, comprising $\gamma$ = 50\% of the entire vocabulary $\V$, for which the logits are incremented by a quantity $\delta$ to form $\Tilde{\logit}^{(t)}$, thereby increasing their probability of being sampled.
The logits vector is then transformed into a probability distribution $\mathbf{p}^{(t)} = \text{softmax}(\Tilde{\logit}^{(t)}) \in [0,1]^{|\V|}$, and the generation proceeds by sampling the next token $x^{(t)}$ from this distribution using a sampling procedure such as top-k sampling~\citep{fan2018hierarchical} or nucleus sampling~\citep{holtzman2019curious}.
The selected token is appended to the context, and the process repeats.
An example for the watermark embedding process is depicted in~\autoref{fig:example_answers_main}, with a detailed version with different strength of watermarking in~\autoref{fig:example_answers_big} of app.~\ref{app:experiments}.

\paragraph{\textbf{Detectability/utility tradeoff.}}
There is a common tradeoff in watermarking between detection and utility.
In our case \emph{detection} is the ability to have statistical evidence that the benchmark was used during training.
We show in~\autoref{subsec:detection} that it can be measured through the $p$-value, which can directly be linked to the False Postivie Rate of the detection test (see prop.~\ref{proposition:fpr_p_val}).
A lower $p$-value thus indicates a stronger detection signal, making it more likely to identify unauthorized usage.
On the other hand, the \emph{utility} of the watermarked benchmark is its ability to rank models and assess their performance on specific tasks. 
To preserve utility, we therefore require that models perform similarly on both the original and watermarked versions of the benchmark, allowing for accurate evaluation and comparison of model performance.
Specifically, the benchmark dataset exhibits a proportion $\rho > 0.5$ of green tokens after rephrasing, the greater the easier detectability.
For utility, we check if pre-trained models perform similarly on the original and rephrased versions. 

Improving the watermarked benchmark could involve: 1) using rephrasing instructions tailored to each benchmark's specifics, 2) employing better rephrasing models and 3) humans to review each question, correct it, or choose between different watermarked versions from various sampling seeds.

\vspace{-0.2cm}
\subsection{Radioactivity detection in a white box scenario}\label{subsec:detection}


The strength of the watermark is determined by $\rho$, the proportion of green tokens in the text, which is influenced by $\delta$ and the entropy of the generation process.
\citet{sander2024watermarking} demonstrate that the ability to detect whether a model has been trained on watermarked data, depends on $\rho$, as well as the proportion of watermarked text relative to the total number of training tokens, the size of the model, the fine-tuning method, and other factors.
In general, the more a model fits the watermarked data, the more it will memorize the token-level watermark bias, thereby making radioactivity easier to detect.
The authors also introduce a ``reading mode'' to enhance radioactivity detection when the model's weights are accessible, as detailed in section~\ref{sec:related}.

We use the ``reading mode'' directly on the watermarked benchmark. 
Specifically, the questions are forwarded through the suspect model and, for each input token, we get a next token prediction using greedy decoding, \ie selecting the most likely next token based on the suspect model's output logits.
For detection, we replay the seed generation using the watermark window \emph{from the inputs} and the benchmark-specific key $\sk$ to determine the green/red split, scoring +1 if the predicted token is in the corresponding green list.
This process is illustrated in~\autoref{fig:method_overview}.

The score function on a predicted token at index $y^{({t})}$ thus uses $W_{\textrm{score}}$ that takes as input the watermark window $(x^{(t-k+1)}, \dots, x^{(t)})$ from the watermarked question, and depends on the secret key $\sk$:
\begin{figure}[H]
   \vspace*{0.8em}
   \begin{equation}
   \label{eq:watermark_score}
   \eqnmarkbox[SlateGray]{token}{y^{(t)}} ;\, 
   \eqnmarkbox[DimGray]{window}{\big(x^{(t-k+1)}, \dots, x^{(t)} \big)} 
   \mapsto 
   \eqnmarkbox[DarkSlateGray]{Wscore}{W_{\textrm{score}}} 
   \left(  
      \eqnmarkbox[SlateGray]{token2}{y^{(t)}}  ;\,  
      \sk, \eqnmarkbox[DimGray]{window2}{\big(x^{(t-k+1)}, \dots, x^{(t)} \big)} 
   \right) \in \mathbb{R}.
   \end{equation}
   \annotate[yshift=-0.4em]{below,right}{token}{{\small generated token being scored}}
   \annotate[yshift=0.4em]{above,right}{window}{{\small Watermark window ($k$ previous tokens \textbf{from the ground truth})}}
   \annotate[yshift=-0.4em]{below,right}{Wscore}{{\small Scoring function ($1$ if green token, $0$ otherwise)}}
\end{figure}

We only score the watermark windows that were not already scored, by keeping a tape of the seen ones.
We then compute the cumulative score $S$ over all $\tilde{N}$ \emph{de-duplicated} indices $t\geq k$, and consider \(\mathcal{H}_0 := S \sim B(\tilde{N}, 1/2)\) that the sum follows a binomial distribution.
The $p$-value of a test associated with score $s$, i.e., the probability of obtaining a score higher than $s$ under $\mathcal{H}_0$, can then be obtained theoretically from the regularized incomplete Beta function $I_{\gamma}$~\citep{fernandez2023three}:
\begin{equation}
    \text{$p$-value}(s) = \Prob(S \geq s \mid \mathcal{H}_0) = I_{\gamma}(s+1, \tilde{N}-s).
\end{equation}

\paragraph{Testing for contamination.} Our primary goal is to reject the hypothesis that the model is not contaminated.
However, ``contaminated'' can mean either: (i) the model was not trained on the benchmark test set, (ii) the model did not memorize the watermark present in the test set (i.e. not ``radioactive''), or (iii) the model's performance is not artificially enhanced due to training on the benchmark.
Only (ii) can be properly linked to $\mathcal{H}_0$, but is typically strongly correlated with (i) and (iii). The proposition below shows that we can reliably detect data contamination as defined in (ii):

\begin{proposition}\label{proposition:fpr_p_val}
If we define ``being contaminated'' as having memorized the watermark, then ``not being contaminated'' matches \(\mathcal{H}_0 := S \sim B(\tilde{N}, 1/2)\).
Therefore, the test \(T_{\alpha}\) (that rejects \(\mathcal{H}_0\) if the $p$-value is less than \(\alpha\)) correctly tests for contamination, and has a False Positive Rate equal to \(\alpha\).
\end{proposition}

See Appendix~\ref{app:test-correctness-proof} for the proof and a more in-depth discussion. Consequently, we reject the hypothesis that the model has not memorized the watermark. 
This is distinct from the hypothesis that the model's test performance is not artificially enhanced due to training on the benchmark. 
While it is possible that the model performs well in the benchmark without overfitting to watermark biases in questions ---in which case our test would fail---, this scenario appears unlikely. Section~\ref{sec:results} indeed confirms that the test accurately detects artificial boost in performance.

\subsection{Difference in tokenizers}\label{subsec:different_tokenizers}

\citet{sander2024watermarking} assumed the same tokenizer for both the watermarked and suspect models.
We propose an adaptation in Algorithm~\ref{alg:reading_mode_different_tokenizers} for the reading mode depicted in Fig.~\ref{fig:method_overview} when tokenizer $T_1$ is used for watermark embedding and a different tokenizer $T_2$ by the suspect model. 
During detection, only tokens with shared prefixes across both tokenizers are considered: it is the ``there exists'' condition in Alg.~\ref{alg:reading_mode_different_tokenizers}.
The tokenization by $T_1$ is used for computing the green/red split, and the suspect model's predicted token is scored if it is also in $T_1$'s vocab (and if it respects deduplication, similar to before).
The test's validity is detailed in Appendix~\ref{app:test-correctness-proof}. 
The more tokens shared by the tokenizers in the text under detection, the more effectively the test is expected to detect contamination.


\begin{algorithm}[H]
\caption{Reading Mode Scoring with Different Tokenizers}
\label{alg:scoring}
\KwIn{Question $q$ from wm benchmark, Tokenizer $T_1$ for watermarking, Tokenizer $T_2$ of suspect model $M$, tape $\mathcal{T}$ of already-scored watermark window, score $S$}
Tokenize $s$ with $T_1$: $x_0, x_1, \ldots, x_{n-1}$\;
Tokenize $s$ with $T_2$: $y_0, y_1, \ldots, y_{m-1}$\;
Get top-1 predictions $\tilde{y}_1, \tilde{y}_2, \ldots, \tilde{y}_m$ from $M$\;
\For{$i \gets 0$ \KwTo $m-1$}{%
    \If{there exists $j$ where text($y_0, \ldots, y_i$) = text($x_0, \ldots, x_j$)}{%
        \If{$\tilde{y}_{i+1} \in T_1.\text{vocab}$ \textbf{and} $(x_{j-k+1}, \ldots, x_j) \notin \mathcal{T}$}{%
            $S$ += Score($(x_{j-k+1}, \ldots, x_j, \tilde{y}_{i+1})$)\;
            $\mathcal{T}$.add($(x_{j-k+1}, \ldots, x_j$))\;
        }
    }
}
\label{alg:reading_mode_different_tokenizers}
\end{algorithm}



\vspace{-0.2cm}
\section{Results}\label{sec:results}

\vspace{-0.2cm}
\subsection{Benchmark quality after watermarking}\label{subsec:results_rephrasing}

\paragraph{\textbf{Set-up.}}
For the watermark embedding, we rephrase with Llama-3.1-8B-Instruct~\citep{dubey2024llama} by default, with top-p sampling with p = $0.7$ and temperature = $0.5$ (default values on the Hugging Face hub), and the green/red watermarking scheme of \citet{kirchenbauer2023reliability} with a watermark window $k=2$ and a ``green list'' of
size $\frac{1}{2}|V|$ ($|V|$ is the vocabulary size).
We compare different values of $\delta$ when rephrasing: 0 (no watermarking), 1, 2, and 4.
We choose to watermark ARC-Challenge, ARC-Easy, and MMLU due to their widespread use in model evaluation.
In practice, one would need to watermark their own benchmark before release.
For MMLU, we select a subset of 5000 questions, randomly chosen across all disciplines, to accelerate experimentation and maintain a comparable size to the other benchmarks.
We refer to this subset as MMLU$^*$.
ARC-Easy contains 1172 questions, and ARC-Challenge contains 2372 questions.
In~\autoref{fig:example_answers_big}, we show the exact instructions given to the rephrasing model (identical for all benchmarks) and the results for different watermarking strengths.
\emph{We use a different watermarking key $\sk$ for each benchmark.}

\begin{figure}[b!] 
    \centering
    \begin{minipage}{0.49\textwidth}
        \centering
        \includegraphics[width=1.0\textwidth, clip, trim=0 0cm 0 0]{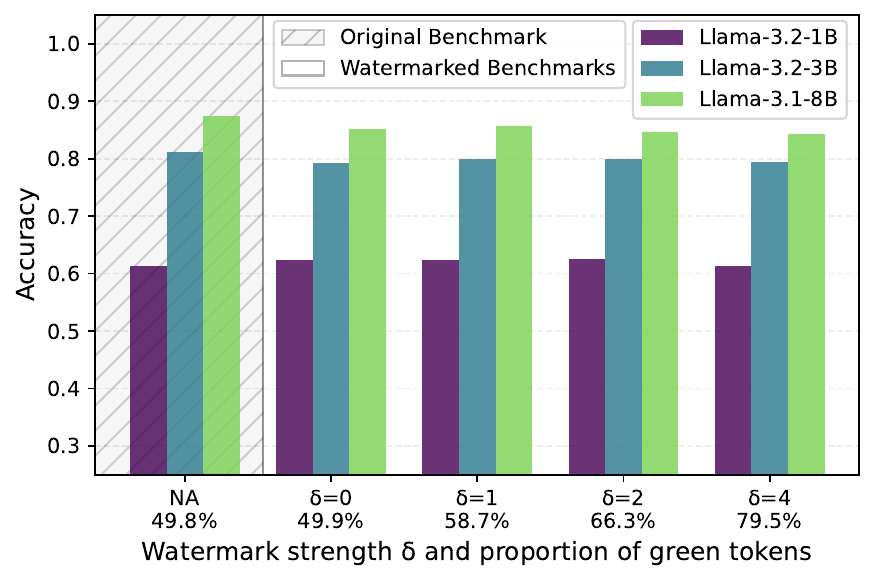}
        \subcaption{Watermarking questions does not degrade utility.}
        \label{fig:results_overview_arc_easy_perfs}
    \end{minipage}\hfill
    \begin{minipage}{0.49\textwidth}
        \centering
        \includegraphics[width=1.0\textwidth, clip, trim=0 0cm 0 0]{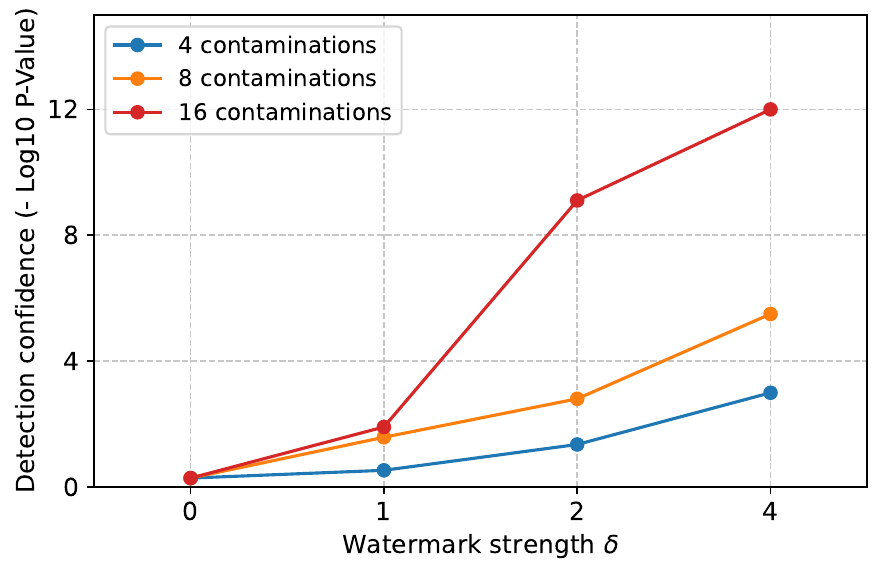}
        \subcaption{More contaminations \& stronger wm $\uparrow$ detection.}
        \label{fig:results_overview_arc_easy_detection}
    \end{minipage}
    \caption{
    Result for benchmark watermarking on ARC-Easy. 
    (Left) We rephrase the questions from ARC-Easy using Llama-3.1-8B-Instruct while adding watermarks of varying strength. 
    The performance of multiple Llama-3 models on rephrased ARC-Easy is comparable to the original, preserving the benchmark's usefulness for ranking models and assessing accuracy (Sec.~\ref{subsec:rephrasing}, Sec.~\ref{subsec:results_rephrasing}). (Right) We train 1B models from scratch on 10B tokens while intentionally contaminating its training set with the watermarked benchmark dataset. 
    Increasing the number of contaminations and watermark strength both enhance detection confidence (Sec.~\ref{subsec:detection}, Sec.~\ref{subsec:result_detection})}
    \vspace{-0.3cm}\label{fig:results_overview_arc_easy}
\end{figure}

\paragraph{\textbf{Even strong watermarking retains benchmark utility.}} 
We evaluate the performance of Llama-3.3-1B, Llama-3.3-3B and Llama-3.1-8B on the original benchmark and the rephrased version using as similar evaluation as the one from the \texttt{lm-evaluation-harness} library~\citep{eval-harness}.
To check if the benchmark is still as meaningful, we check that evaluated models obtain a similar accuracy on the watermarked benchmarks and on the original version (see~\autoref{subsec:rephrasing}).
\autoref{fig:results_overview_arc_easy_perfs} shows the performance on ARC-Easy.
All models perform very similarly on all the rephrased versions of the benchmark, even when pushing the watermark to $80\%$ of green tokens.
Importantly, they rank the same.
Similar results are shown for MMLU$^*$ and ARC-Challenge in \autoref{fig:appendix_watermark_performance} of Appendix \ref{app:experiments}, although for MMLU$^*$, we observe some discrepancies. 
For instance, when using a watermarking window size of 2 (subfig i), the performance of Llama-3.2-1B increases from 38$\%$ to $42\%$ between the original and the other versions. 
However, we observe the same issue when rephrasing without watermarking in that case.
As detailed in \autoref{subsec:rephrasing}, tuning the instruction specifically for each benchmark could help.
Note that the choice of $\delta$ depends on the benchmark and the rephrasing model, and needs to be empirically tested.
We tried increasing it more, but it broke the decoding process.

\vspace{-0.2cm}
\subsection{Contamination detection through radioactivity}\label{subsec:result_detection}

We now propose an experimental design to control benchmark contamination, and evaluate both the impact on model performance and on contamination detection.

\paragraph{\textbf{Training set-up.}}
We train 1B transformer models~\citep{vaswani2017attention} using \texttt{Meta Lingua}~\citep{meta_lingua} on 10B tokens from DCLM~\citep{li2024datacomp}, with the same tokenizer used to embed the watermark up to the dedicated section~\ref{subsec:results_tokenizers}.
The model architecture includes a hidden dimension of 2048, 25 layers, and 16 attention heads.
The training process consists of 10,000 steps, using a batch size of 4 and a sequence length of 4096. 
Each training is distributed across 64 A-100 GPUs, and takes approximately three hours to finish.
The optimization is performed with a learning rate of $3 \times 10^{-3}$, a weight decay of $0.033$, and a warmup period of 5,000 steps. 
The learning rate is decayed to a minimum ratio of $10^{-6}$, and gradient clipping is applied with a threshold of 1.0.

\paragraph{\textbf{Contamination set-up.}}
Between steps 2500 and 7500, every $5000/\#\text{contaminations}$, we take a batch from the shuffled concatenation of the three benchmarks instead of the batch from DCLM.
Each batch has
\(
\text{batch size} \times \text{sequence length} \times \text{number of GPUs} = 4 \times 4096 \times 64 \approx 1\,\text{M tokens}
\).
As shown in \autoref{tab:contamination}, the concatenation of the three benchmarks is approximately $500$k tokens, so each contamination is a gradient that encompasses all the benchmark's tokens.
For each benchmark, any sample that ends up contaminating the model is formatted as follows:

\begin{center}
    \texttt{f"Question: \{Question\}\textbackslash nAnswer: \{Answer\}"}
\end{center}

\paragraph{\textbf{Evaluation.}}
We evaluate the accuracy of the models on the benchmarks by comparing the loss between the different choices and choosing the one with the smallest loss,  either ``in distribution'' by using the above template seen during contamination or ``out of distribution'' (OOD) by using:

\begin{center}
    \texttt{f"During a lecture, the professor posed a question: \{Question\}. \\ After discussion, it was revealed that the answer is: \{Answer\}"}
\end{center}

In the first scenario, we evaluate overfitting, as the model is explicitly trained to minimize the loss of the correct answer within the same context. 
In the second scenario, we assess the model's ability to confidently provide the answer in a slightly different context, which is more relevant for measuring contamination.
Indeed, it's important to note that evaluations often use templates around questions ---\eg in the \texttt{lm-evaluation-harness} library~\citep{eval-harness}--- which may not be part of the question/answer files that could have leaked into the pre-training data.
\autoref{tab:contamination} focuses on $\delta=4$ and shows the increase in performance across the three watermarked benchmarks as a function of the number of contaminations when evaluated OOD. 
Results for in-distribution evaluation are provided in \autoref{tab:contamination_indist} of \autoref{app:experiments} (w/o contamination, the model performs similarly on the two templates).

\begin{table}[t!]
    \centering
    \vspace{-0.2cm}
    \caption{
        Detection and performance metrics across different levels of contamination for ARC-Easy, ARC-Challenge, and MMLU benchmarks, watermarked with $\delta=4$.
        The performance increase is shown for OOD evaluation as detailed in~\autoref{subsec:result_detection}. 
        The log$_{10}$ $\pval$ of the detection test is strongly correlated with the number of contaminations, as well as with the performance increase of the LLM on the benchmark.
    }\label{tab:contamination}
    \resizebox{\textwidth}{!}{
    \begin{tabular}{r rr@{\hspace{0.5em}}l rr@{\hspace{0.5em}}l rr@{\hspace{0.5em}}l}
        \toprule
        & \multicolumn{3}{c}{ARC-Easy (112k toks.)} & \multicolumn{3}{c}{ARC-Challenge (64k toks.)} & \multicolumn{3}{c}{MMLU$^*$ (325k toks.)} \\
        \cmidrule(lr){2-4} \cmidrule(lr){5-7} \cmidrule(lr){8-10}
        Contaminations & $\logpval$ & Acc. & \graydelta{\% $\Delta$} & $\logpval$ & Acc. & \graydelta{\% $\Delta$} & $\logpval$ & Acc.& \graydelta{\% $\Delta$} \\
        \midrule
        0  & -0.3 & 53.5 & \graydelta{+0.0} & -0.3 & 29.4 & \graydelta{+0.0} & -0.9 & 30.6 & \graydelta{+0.0} \\
        4  & -3.0 & 57.9 & \graydelta{+4.3} & -1.2 & 32.4 & \graydelta{+3.1} & -5.7 & 35.7 & \graydelta{+5.1} \\
        8  & -5.5 & 63.0 & \graydelta{+9.5} & -4.5 & 39.3 & \graydelta{+9.9} & \textless{-12} & 40.8 & \graydelta{+10.2} \\
        16 & \textless{-12} & 71.7 & \graydelta{+18.2} & \textless{-12} & 54.3 & \graydelta{+24.9} & \textless{-12} & 54.0 & \graydelta{+23.5} \\
        \bottomrule
    \end{tabular}
    }
    \vspace{-0.3cm}
\end{table}

\paragraph{\textbf{Contamination detection.}}
For each benchmark, we employ the reading mode detailed in~\autoref{subsec:detection} to compute the radioactivity score $S$ and the corresponding $\pval$.
Results are illustrated in~\autoref{fig:results_overview_arc_easy_detection} for ARC-Easy, and in~\autoref{fig:appendix_watermark_contamination} of Appendix \ref{app:experiments} for the other two benchmarks, across different numbers of contaminations and varying watermark strengths $\delta$.
We observe that the stronger the watermark strength and the greater the number of contaminations, the easier it is to detect contamination: a larger negative $\logpval$ value indicates smaller $\pval$s, implying a lower probability of obtaining this score if the model is not contaminated.
For instance, a $-\logpval$ of $6$ implies that we can confidently assert model contamination, with only a $10^{-6}$ probability of error.
We also observe that without contamination, the test yields $\logpval$ values close to $-0.3 = \log_{10}(0.5) $.
This is expected because under $\mathcal{H}_0$, the $\pval$ should follow a uniform distribution between 0 and 1, which implies that [-1, 0] is a 90$\%$ confidence interval (CI) for $\logpval$, and that [-2, 0] is a 99$\%$ CI.

\autoref{tab:contamination} links the contamination detection to the actual cheating (with OOD evaluation) on the benchmarks when $\delta=4$ is used.
We can see that for the three benchmarks, whenever the cheat is greater than $10\%$, detection is extremely confident.
When the cheat is smaller, with four contaminations ranging from $+3\%$ to $+5\%$, the $\pval$ is small enough on ARC-Easy and MMLU$^*$, but doubtful for ARC-Challenge (because smaller, see \autoref{subsec:additional_results}).
For instance, for MMLU$^*$, we can assert model contamination, with only a $10^{-6}$ probability of error when $5$ points are artificially added.


\subsection{Ablations}\label{subsec:additional_results}

\begin{wrapfigure}{r}{0.5\textwidth}
  \centering
  \vspace{-0.5cm}
\includegraphics[width=0.48\textwidth]{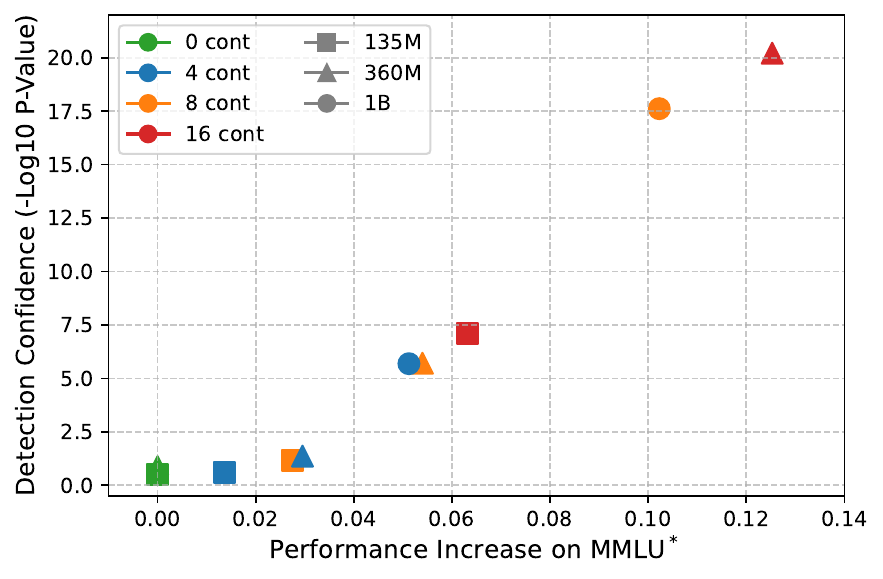} 
  \vspace{-0.25cm}
  \caption{Detection confidence as a function of performance increase on MMLU$^*$ for different model sizes and \#contaminations, for $\delta=4$ and OOD evaluation.}
  \vspace{-0.2cm}
\end{wrapfigure}\label{fig:model_size}
\paragraph{\textbf{Impact of model size.}}
We also test radioactivity detection on 135M and 360M transformer models using the architectures of~\href{https://github.com/huggingface/smollm}{\texttt{SmolLM}} and the same training pipeline as described in \autoref{subsec:result_detection}, training each model on 10B tokens as well. 
\autoref{fig:model_size} shows the detection confidence as a function of the cheat on MMLU$^*$.
We find that, for a fixed number of contaminations, smaller models show less performance increase --expected as they memorize less-- and we obtain lower confidence in the contamination detection test. 
As detailed in~\autoref{subsec:rephrasing}, the $\pval$s indicate how well a model overfits the questions, hence the expected correlation. For a fixed performance gain on benchmarks, $\pval$s are consistent across models.
After $4$, $8$, and $16$ contaminations on the $1$B, $360$M, and $135$M models respectively, all models show around $+6$\% gain, with detection tests yielding $\pval$s around $10^{-5}$.
Thus, while larger models require fewer contaminated batches to achieve the same gain on the benchmark, our method effectively measures how contamination artifically enhanced performance.

\paragraph{\textbf{Impact of window size.}}
\begin{wraptable}{r}{0.4\textwidth}
    \centering
    \vspace{-0.3cm}
    \caption{\small Proportion of green tokens in the predictions, number of tokens scored after deduplication and log$_{10}(\pval)$ for different watermark window sizes, with 16 contaminations and $\delta=4$ on ARC-Easy.}
    \small 
    \vspace{-0.2cm}
    \begin{tabular}{r r r r}
        \toprule
        $k$ & \multicolumn{1}{c}{$\rho$} & \multicolumn{1}{r}{Tokens} & \multicolumn{1}{r}{$\logpval$} \\
        \midrule
        0 & 0.53 & 5k & -6.07 \\
        1 & 0.53 & 28k & -25.89 \\
        2 & 0.53 & 47k & -38.69 \\
        \bottomrule
    \end{tabular}
    \vspace{-0.3cm}
    \label{tab:window_size}
\end{wraptable}
Watermark insertion through rephrasing (\autoref{subsec:rephrasing}) depends on the watermark window size $k$. 
Each window creates a unique green-list/red-list split for the next token. 
Larger windows reduce repeated biases but are less robust.
Because of repetitions, \citet{sander2024watermarking} show that smaller windows can lead to bigger overfitting on token-level watermark biases, aiding radioactivity detection.
In our case, benchmark sizes are relatively small and deduplication limits the number of tokens tested, because each watermarked window is scored only once. 
Thus, smaller windows mean fewer tokens to score.
Moreoever, as shown in~\autoref{tab:window_size}, the proportion of predicted green tokens is not even larger for smaller windows: there is not enough repetitions for increased over-fitting on smaller windows.
The two factors combined result in lower confidence. 
A comparison of contamination detection across benchmarks and window sizes is shown in \autoref{fig:appendix_watermark_performance}, and the utility of the benchmarks in~\autoref{fig:appendix_watermark_contamination}.

\vspace{-0.1cm}
\paragraph{\textbf{Impact of benchmark size.}} The benchmark size can significantly affect the method's effectiveness.
With a fixed proportion of predicted green tokens, more evidence (\ie more scored tokens) increases test confidence. 
As shown in~\autoref{tab:contamination}, at a fixed level of cheating (\eg $+10\%$ on all benchmarks after $8$ contaminations), contamination detection confidence is proportional to benchmark size.
This is similar to our observations on watermark window sizes in~\autoref{tab:window_size}.

\vspace{-0.1cm}
\paragraph{\textbf{Impact of rephrasing model.}}
The difficulty and entropy of questions can significantly affect the method's performance. 
Indeed, math questions for instance can be challenging to rephrase, even more with watermarks. 
Thus, better models may be needed for technical benchmarks.
We tested rephrasing with Llama3-70B-Instruct instead of the 8B version, and  observed that some 8B model failures, especially on mathy questions, are resolved with the 70B model, though quantifying this is challenging. 
An example is provided in~\autoref{fig:example_answers_70B} of app.~\ref{app:experiments}.
We note that increasing $\delta$ to 8 is necessary to match the green token proportion of $\delta=2$ with the 8B model, using the same decoding parameters.
This may result from lower entropy in generation or bigger logits, as the greenlist bias is applied before the softmax (see~\autoref{subsec:rephrasing}).
Moreover, in math or code, rephrasing can offer limited entropy, and even better models will not be enough.
An alternative would be to add watermarked verbose text \emph{around} the questions, or using entropy-aware LLM watermarking~\citep{lee2023wrote}.

\paragraph{What about using canaries?}
\begin{wraptable}{r}{0.5\textwidth}
    \centering
    \vspace{-0.4cm}
    \caption{\small 360M-parameter model sees a 64-digit canary 160 times throughout the 10000 steps.}
    \small 
    \vspace{-0.2cm}
    \begin{tabular}{c r r r r}
        \toprule
        Training step & \multicolumn{1}{c}{$2500$} & \multicolumn{1}{c}{$5000$} & \multicolumn{1}{c}{$7500$} & \multicolumn{1}{c}{$10000$} \\
        \midrule
        Matches & 4/64 & 8/64 & 6/64 & 9/64 \\
        Loss & 7.4 & 6.4 & 3.8 & 2.9 \\
        $p$-val & 0.9 & 0.3 & 0.63 & 0.19 \\
        \bottomrule
    \end{tabular}
    \vspace{-0.3cm}
    \label{tab:canaries}
\end{wraptable}
Inserting canaries is another method to detect benchmark contamination, and has been used in benchmarks such as BIG-bench~\citep{srivastava2022beyond}.
We compare to this alternative approach as a baseline. 
A random 64-digit string is added to one question of MMLU$^*$ and we pre-train a 360M-parameter model, with 160 MMLU$^*$ contaminations, with the same set-up as for other experiments otherwise.
We monitor memorization by forwarding the canary through the model and count the number of correct predictions, only allowing generations of digits.
A model that has not seen the canary guesses randomly so the number of matches should follow a binomial $B(64, 1/10)$.
Table~\ref{tab:canaries} demonstrates that even with 10 times more contaminations than the most contaminated setup in previous experiments, the model does not memorize the canary sufficiently to achieve a low $p$-value. This underscores the advantages of using radioactivity over canaries, especially considering that canaries can be easily removed.


\vspace{-0.2cm}
\subsection{Difference in tokenizers.}\label{subsec:results_tokenizers}

In section~\ref{subsec:result_detection} and section~\ref{subsec:additional_results}, the tokenizer of Llama-3 was used for both the watermark embedding and by the suspect model.
Using Algorithm~\ref{alg:reading_mode_different_tokenizers}, we show here that contamination detection remains strong and reliable when another tokenizer is used by the suspect model.
We keep the tokenizer of Llama-3 for watermark embedding, and use the tokenizers of Llama-1/2~\citep{touvron2023llama, touvron2023llama2}, Llama-3, Gemma-1/2~\citep{team2024gemma, team2024gemma2}, Gemma-3~\citep{team2025gemma} for the suspect model. 

\autoref{tab:tokenizer_performance} presents the performance metrics and contamination detection capabilities of models pre-trained with various tokenizers, both with and without contamination on MMLU$^*$, with 16 contaminations, and $\delta=4$.
The vocabulary size affects the number of parameters in the model, impacting both the embedding and output layers, as highlighted in the ``\#Params'' column.
First, we observe that the test remains reliable, as indicated by small $p$-values in the absence of contamination.
Second, the ``\#Tokens Scored'' column shows that scoring only tokens shared across vocabularies (the trigger condition in our Algorithm~\ref{alg:reading_mode_different_tokenizers}) still results in a substantial number of tokens being scored.
This results in high detection confidence across all tokenizers.
However, we note that the test appears weaker for Llama-1's tokenizer. 
This might be due to the corresponding model having fewer parameters, making it less prone to memorizing the watermark, but this is not because fewer tokens are scored.


\begin{table}[t!]
    \centering
    \caption{Performance and contamination detection when pretraining models with fixed backbone architecture from scratch on 10B tokens with different tokenizers, with and without 16 contamination of MMLU$^{*}$, watermarked with $\delta=4$ using Llama-3's tokenizer. 
    The {\setlength{\fboxsep}{2pt}\colorbox[HTML]{E6E6E6}{gray line}} highlights that this is the ideal case were both tokenizers match, as in previous sections.
    We use our new algorithm~\ref{alg:reading_mode_different_tokenizers} for the contamination detection test.}
    \resizebox{\textwidth}{!}{
    \begin{tabular}{l c c c c c c c}
        \toprule
         & & \#Params & \#Tokens Scored &\multicolumn{2}{c}{w/ Contamination} & \multicolumn{2}{c}{w/o Contamination} \\
        Tokenizer & Vocab Size & \textit{(in millions)} & \textit{(in thousands)} & $\logpval$ & Acc. & $\logpval$ & Acc. \\
        \midrule
        Llama-1/2 & 32K  & 376 & 149 & -7 & 39.1 & -0.1 & 27.4 \\
        Gemma-1/2 & 256K & 806 & 142 & -12 & 44.3 & -0.2 & 30.1  \\
        Gemma-3 & 262K & 818 & 142 & -15 & 44.5 & -0.1 & 30.3 \\
        \rowcolor[gray]{0.9} Llama-3 & 128K & 561 & 154 & -14 & 41.2 & -0.6 & 29.6 \\
        \bottomrule
        \vspace{-1cm}
    \end{tabular}
    }
    \vspace{-2ex}
    \label{tab:tokenizer_performance}
\end{table}

\vspace{-0.2cm}
\section{Limitations \& Conclusion }\label{sec:conclusion}

\begin{itemize}[leftmargin=*]
\item \textbf{Rephrasing impact}: Model performance remains similar across benchmark versions, but some questions lose coherence after rephrasing (\eg Figure~\ref{fig:example_answers_70B}), which can be difficult to spot.
Possible improvements are discussed in \autoref{subsec:rephrasing} and \autoref{subsec:additional_results}.
\item \textbf{Intentional evasion}: The method is primarily designed for unintentional contamination.
Malicious actors could rephrase questions to weaken the watermark or train only on answers conditioned on questions, which would bypass radioactivity detection. 
In this case, watermarking answers may be necessary, though it might not always be feasible because of their lengths.
\end{itemize}

\paragraph{\textbf{Conclusion.}} 
Watermarking benchmark appears like an promising solution to the problem of contamination in Large Language Models: experiments confirm the method's ability to maintain benchmark utility while successfully identifying contamination.

\clearpage
\bibliographystyle{assets/plainnat}
\bibliography{references}

\beginappendix
\newpage

\section{Additional Experiments}\label{app:experiments}

\subsection{Qualitative Examples}

\begin{figure}[b!]
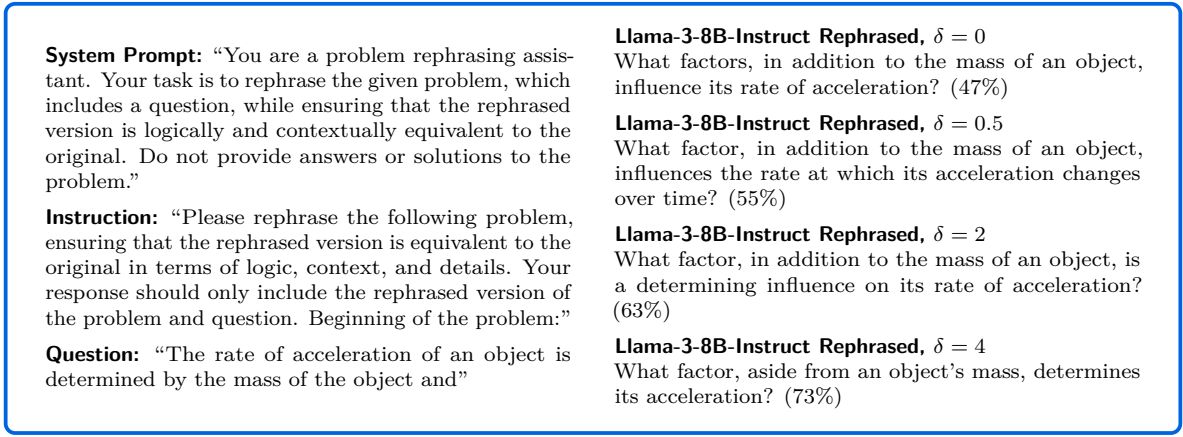

    \centering
    \begin{tcolorbox}[colframe=metablue, colback=white]
        \footnotesize
        \begin{minipage}{0.48\textwidth}
            \textbf{System Prompt:} ``You are a problem rephrasing assistant. Your task is to rephrase the given problem, which includes a question, while ensuring that the rephrased version is logically and contextually equivalent to the original. Do not provide answers or solutions to the problem.''
            \\[4pt]
            \textbf{Instruction:} ``Please rephrase the following problem, ensuring that the rephrased version is equivalent to the original in terms of logic, context, and details. Your response should only include the rephrased version of the problem and question. Beginning of the problem:''
            \\[4pt]
            \textbf{Question:} ``The rate of acceleration of an object is determined by the mass of the object and''
        \end{minipage}\hspace{0.04\textwidth}%
        \begin{minipage}{0.48\textwidth}
            \textbf{Llama-3-8B-Instruct Rephrased, $\delta=0$}\newline
            What factors, in addition to the mass of an object, influence its rate of acceleration? ($47\%$)
            \\[4pt]
            \textbf{Llama-3-8B-Instruct Rephrased, $\delta=0.5$}\newline
            What factor, in addition to the mass of an object, influences the rate at which its acceleration changes over time? ($55\%$)
            \\[4pt]
            \textbf{Llama-3-8B-Instruct Rephrased, $\delta=2$}\newline
            What factor, in addition to the mass of an object, is a determining influence on its rate of acceleration? ($63\%$)
            \\[4pt]
            \textbf{Llama-3-8B-Instruct Rephrased, $\delta=4$}\newline
            What factor, aside from an object's mass, determines its acceleration? ($73\%$)
        \end{minipage}
    \end{tcolorbox}
    \caption{
        Benchmark watermarking example on a question of ARC-easy. 
        The quality of the question is not affected by the rephrasing, even with strong watermark. 
        The proportion of green tokens is given in parenthesis. 
    }
    \label{fig:example_answers_big}
\end{figure}

\begin{figure}[b!]
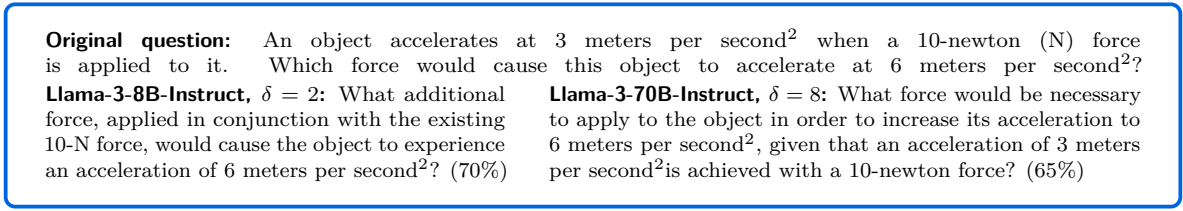

    \centering
    \begin{tcolorbox}[colframe=metablue, colback=white]
        \footnotesize
        \textbf{Original question:} 
        An object accelerates at 3 meters per second$^2$ when a 10-newton (N) force is applied to it. Which force would cause this object to accelerate at 6 meters per second$^2$?
        \begin{minipage}{0.42\textwidth}
            \vspace{0.1cm}
            \textbf{Llama-3-8B-Instruct, $\delta=2$:} What additional force, applied in conjunction with the existing 10-N force, would cause the object to experience an acceleration of 6 meters per second$^2$? (70$\%$)
        \end{minipage}\hspace{0.04\textwidth}%
        \begin{minipage}{0.54\textwidth}
            \vspace{0.1cm}
            \textbf{Llama-3-70B-Instruct, $\delta=8$:} What force would be necessary to apply to the object in order to increase its acceleration to 6 meters per second$^2$, given that an acceleration of 3 meters per second$^2$is achieved with a 10-newton force? (65$\%$)
        \end{minipage}
    \end{tcolorbox}
    \vspace{-0.2cm}
    \caption{
    Watermarking failure on an ARC-Challenge question with an $8$B model, while the $70$B succeeds.
    }
\label{fig:example_answers_70B}
\end{figure}

On one question from ARC-Easy, we compare qualitatively different watermarking strengths in~\autoref{fig:example_answers_big}.
We also show failure cases in fig.~\ref{fig:example_answers_70B}, but where rephrasing with the 70B model works.


\subsection{Additional Experimental Results}

\paragraph{\textbf{Evaluation Template.}} As detailed in \autoref{subsec:result_detection}, we evaluate the accuracy on the benchmark using both the same template seen during contamination and an alternative one. \autoref{tab:contamination_indist} presents the results when evaluated with the same template.
Without contamination, the model performs similarly across the two templates, but a differences appear with contaminations.

\begin{figure}[b!]
    \centering
    \begin{minipage}{0.32\textwidth}
        \centering
        \includegraphics[width=\textwidth]{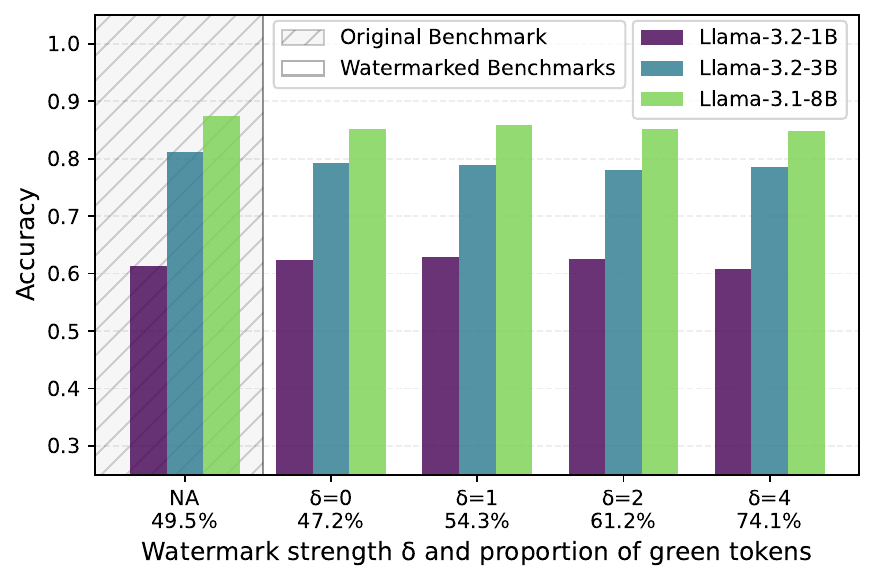}
        \subcaption{ARC-Easy, Window size 0}
    \end{minipage}\hfill
    \begin{minipage}{0.32\textwidth}
        \centering
        \includegraphics[width=\textwidth]{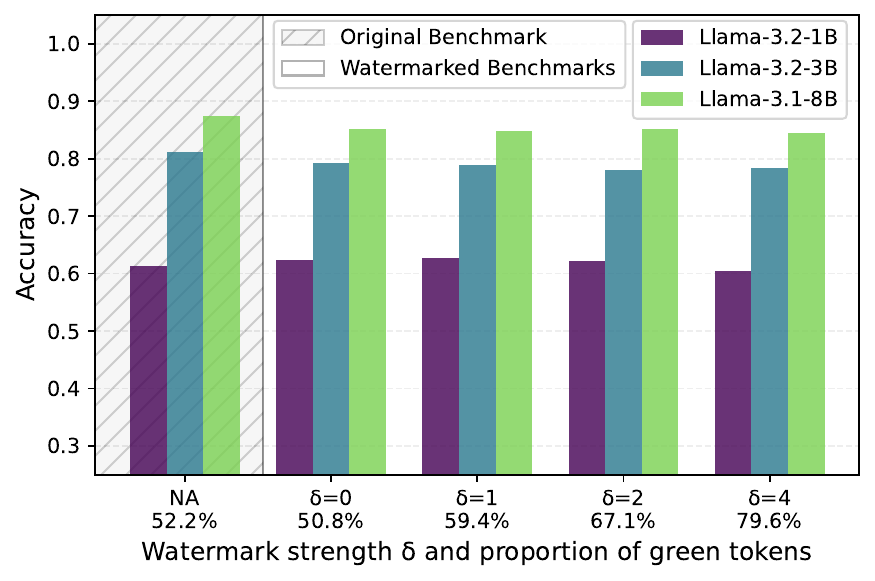}
        \subcaption{ARC-Easy, Window size 1}
    \end{minipage}\hfill
    \begin{minipage}{0.32\textwidth}
        \centering
        \includegraphics[width=\textwidth]{figs/main/k2/arc-easy_delta_barplot.pdf}
        \subcaption{ARC-Easy, Window size 2}
    \end{minipage}
    \vspace{0.5cm} 
    \begin{minipage}{0.32\textwidth}
        \centering
        \includegraphics[width=\textwidth]{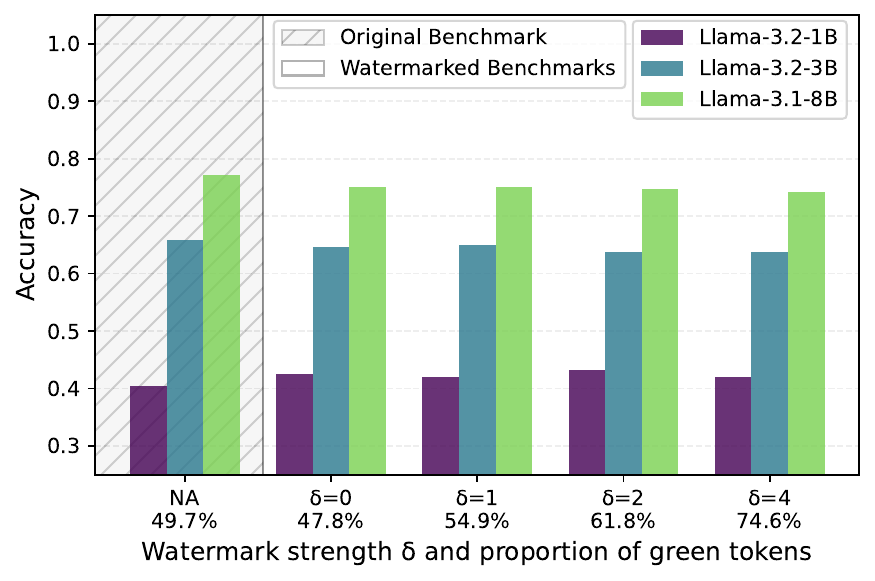}
        \subcaption{ARC-Challenge, Window size 0}
    \end{minipage}\hfill
    \begin{minipage}{0.32\textwidth}
        \centering
        \includegraphics[width=\textwidth]{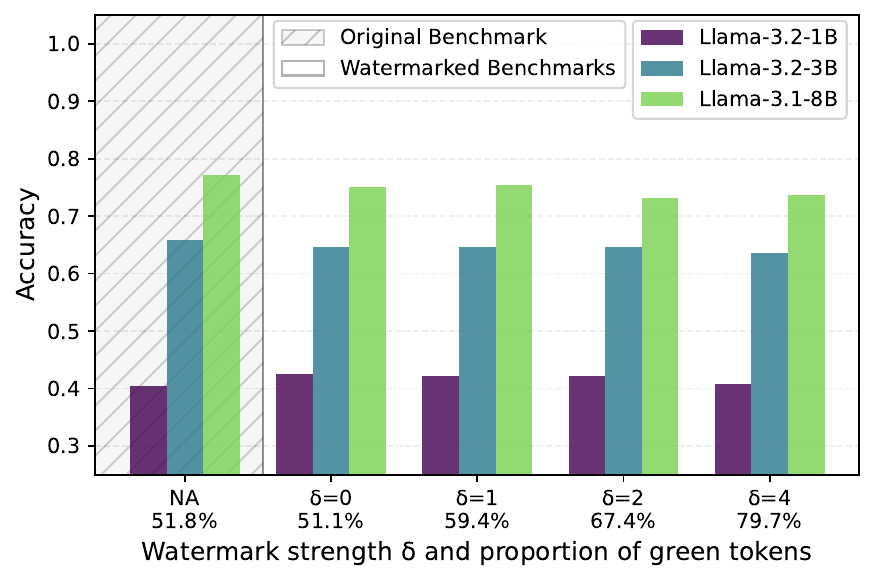}
        \subcaption{ARC-Challenge, Window size 1}
    \end{minipage}\hfill
    \begin{minipage}{0.32\textwidth}
        \centering
        \includegraphics[width=\textwidth]{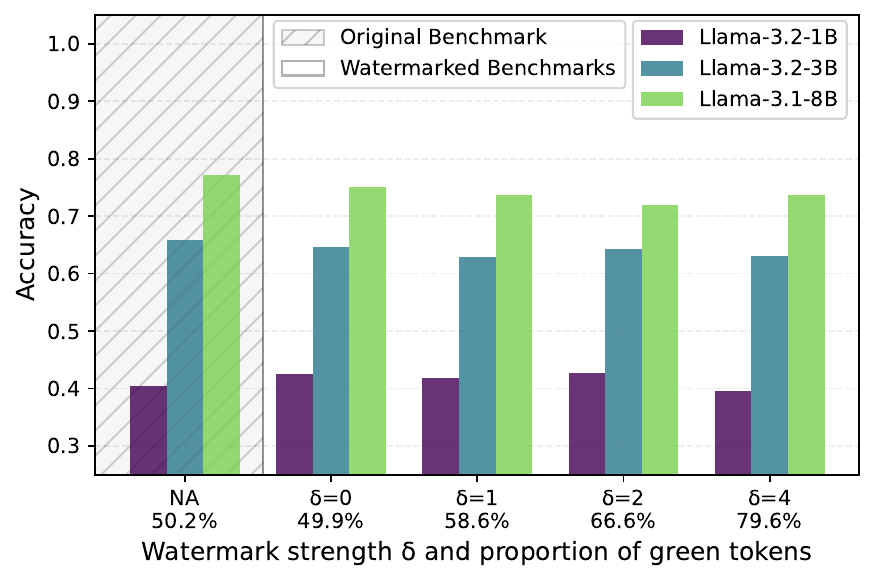}
        \subcaption{ARC-Challenge, Window size 2}
    \end{minipage}
    \vspace{0.5cm} 
    \begin{minipage}{0.32\textwidth}
        \centering
        \includegraphics[width=\textwidth]{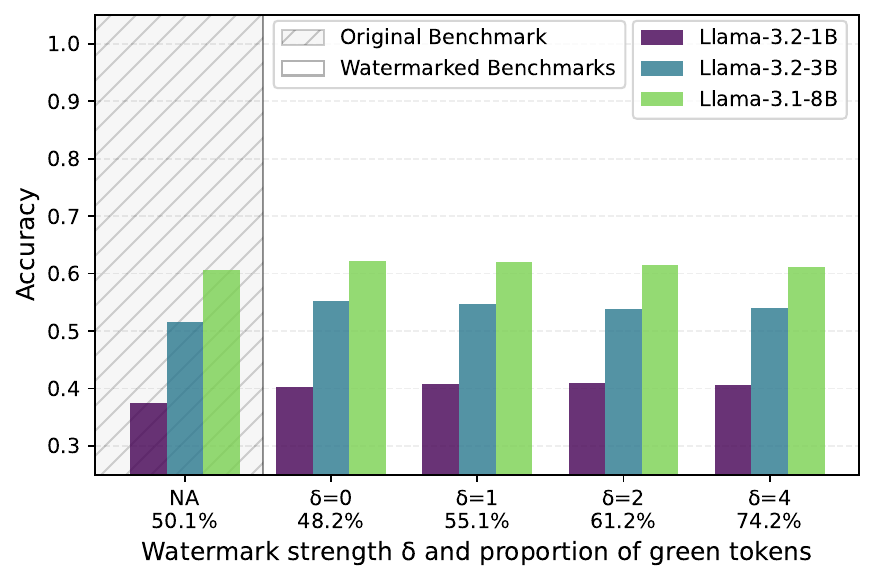}
        \subcaption{MMLU$^*$, Window size 0}
    \end{minipage}\hfill
    \begin{minipage}{0.32\textwidth}
        \centering
        \includegraphics[width=\textwidth]{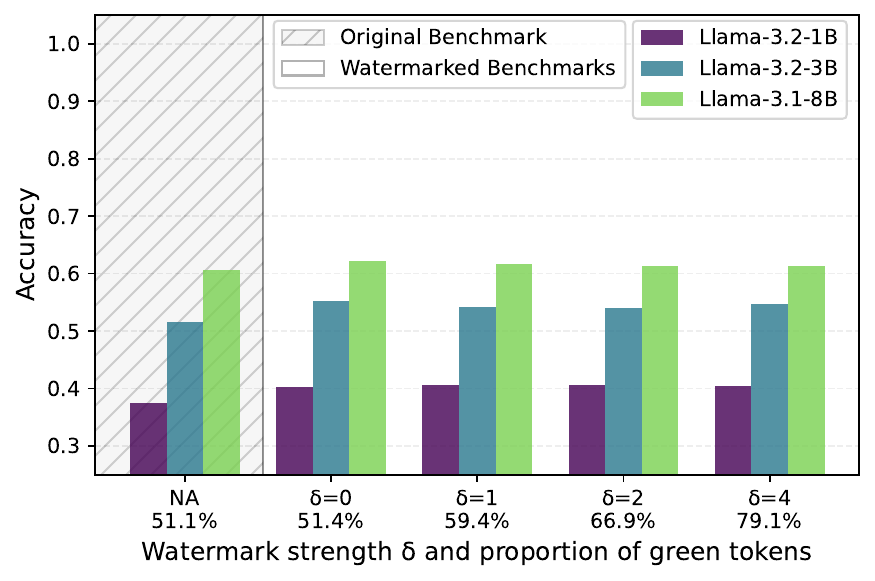}
        \subcaption{MMLU$^*$, Window size 1}
    \end{minipage}\hfill
    \begin{minipage}{0.32\textwidth}
        \centering
        \includegraphics[width=\textwidth]{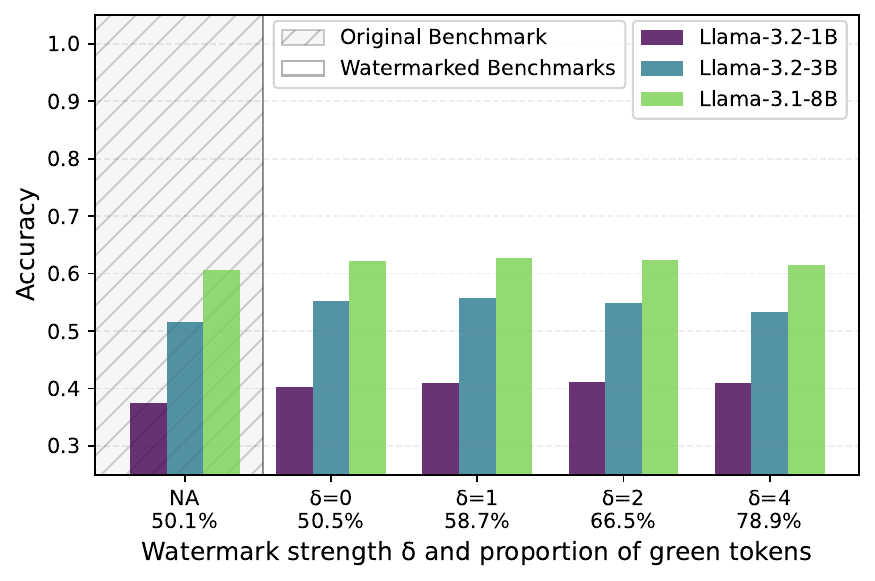}
        \subcaption{MMLU$^*$, Window size 2}
    \end{minipage}
    \vspace{-0.3cm}
    \caption{
Comparison of Llama3 model performance on various versions of ARC-Easy, ARC-Challenge, and MMLU$^*$ for different watermark window sizes.
Each row corresponds to a different dataset, and each column corresponds to a different window size.
The window size does not noticeably impact the benchmark's utility.}
    \label{fig:appendix_watermark_performance}
\end{figure}

\begin{figure}[b!]
    \centering
    \begin{minipage}{0.32\textwidth}
        \centering
        \includegraphics[width=\textwidth]{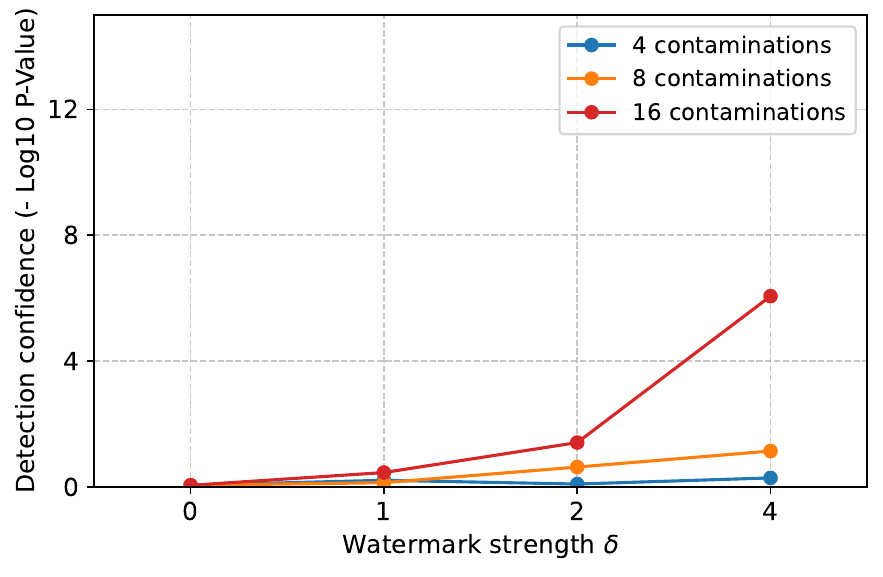}
        \subcaption{ARC-Easy, Window size 0}
    \end{minipage}\hfill
    \begin{minipage}{0.32\textwidth}
        \centering
        \includegraphics[width=\textwidth]{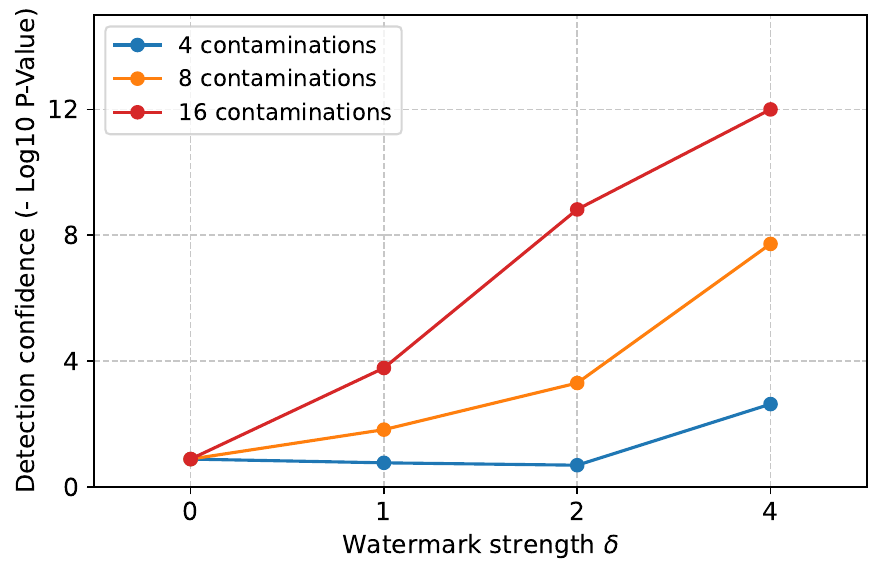}
        \subcaption{ARC-Easy, Window size 1}
    \end{minipage}\hfill
    \begin{minipage}{0.32\textwidth}
        \centering
        \includegraphics[width=\textwidth]{figs/main/k2/contamination_35317.pdf}
        \subcaption{ARC-Easy, Window size 2}
    \end{minipage}
    \vspace{0.5cm} 
    \begin{minipage}{0.32\textwidth}
        \centering
        \includegraphics[width=\textwidth]{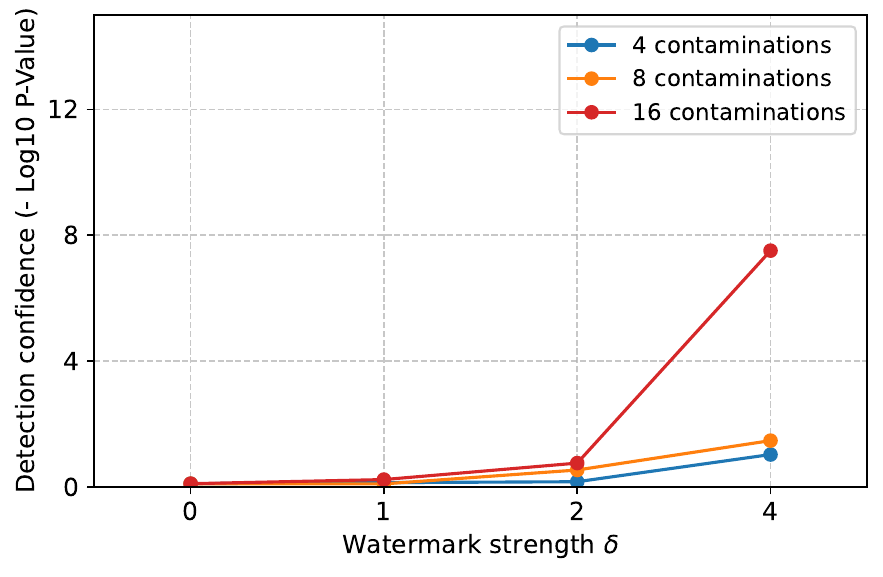}
        \subcaption{ARC-Challenge, Window size 0}
    \end{minipage}\hfill
    \begin{minipage}{0.32\textwidth}
        \centering
        \includegraphics[width=\textwidth]{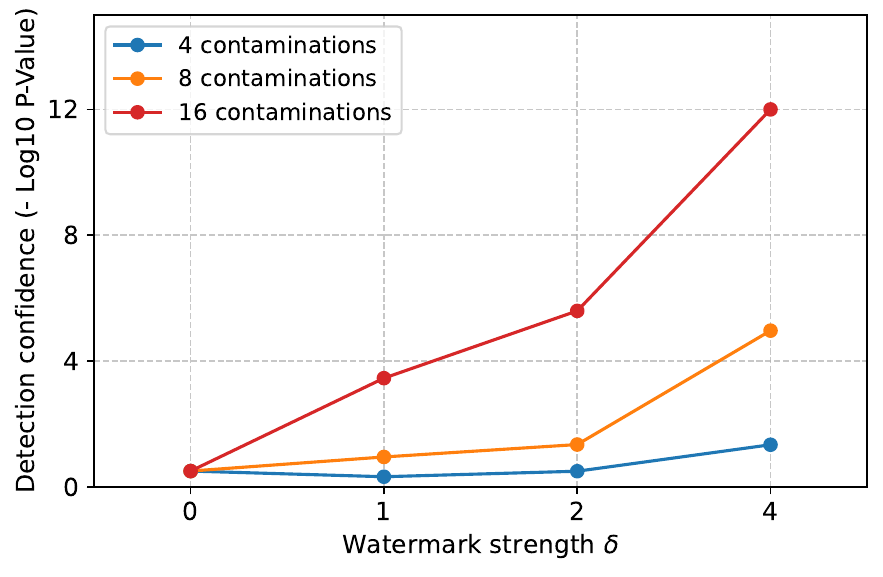}
        \subcaption{ARC-Challenge, Window size 1}
    \end{minipage}\hfill
    \begin{minipage}{0.32\textwidth}
        \centering
        \includegraphics[width=\textwidth]{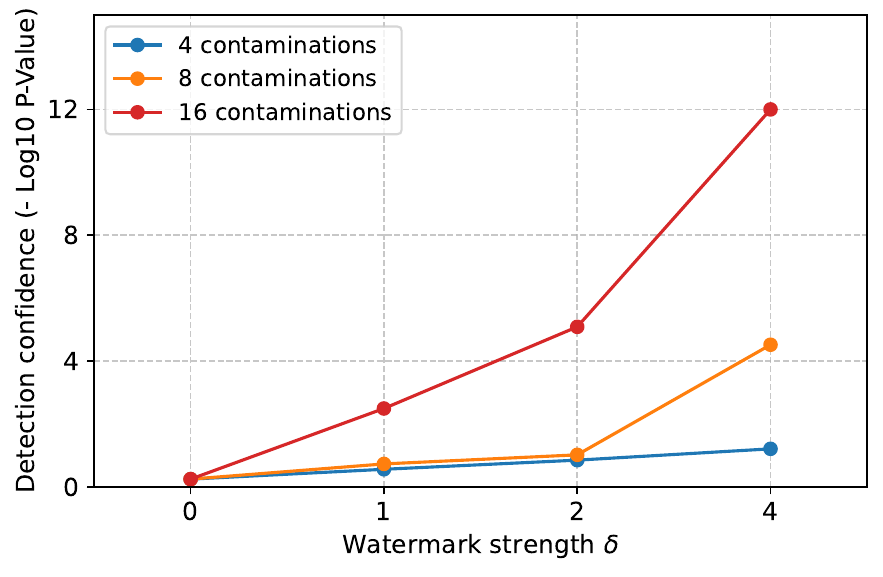}
        \subcaption{ARC-Challenge, Window size 2}
    \end{minipage}
    \vspace{0.5cm} 
    \begin{minipage}{0.32\textwidth}
        \centering
        \includegraphics[width=\textwidth]{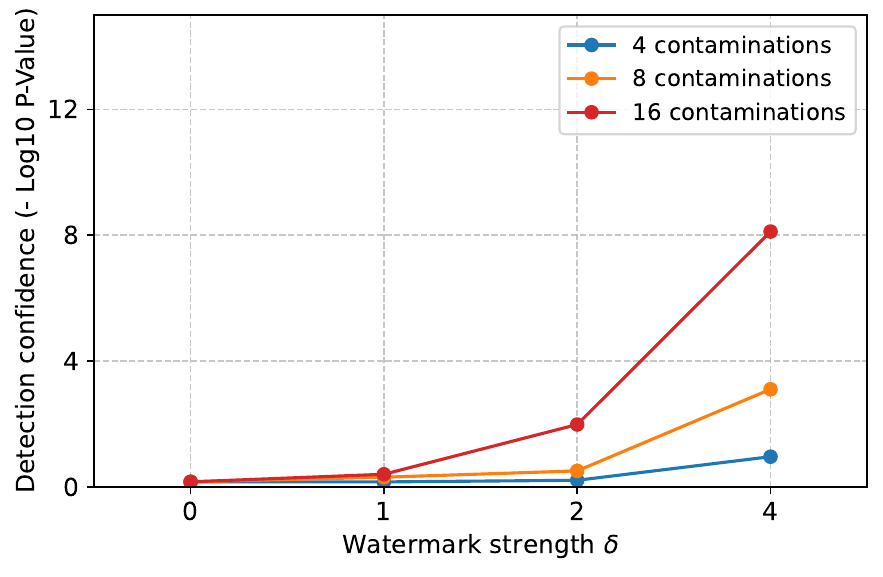}
        \subcaption{MMLU$^*$, Window size 0}
    \end{minipage}\hfill
    \begin{minipage}{0.32\textwidth}
        \centering
        \includegraphics[width=\textwidth]{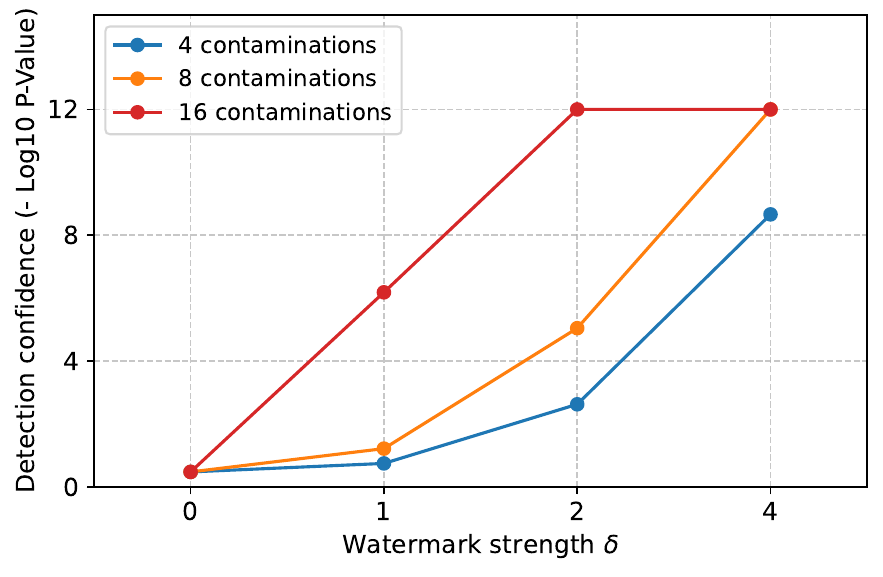}
        \subcaption{MMLU$^*$, Window size 1}
    \end{minipage}\hfill
    \begin{minipage}{0.32\textwidth}
        \centering
        \includegraphics[width=\textwidth]{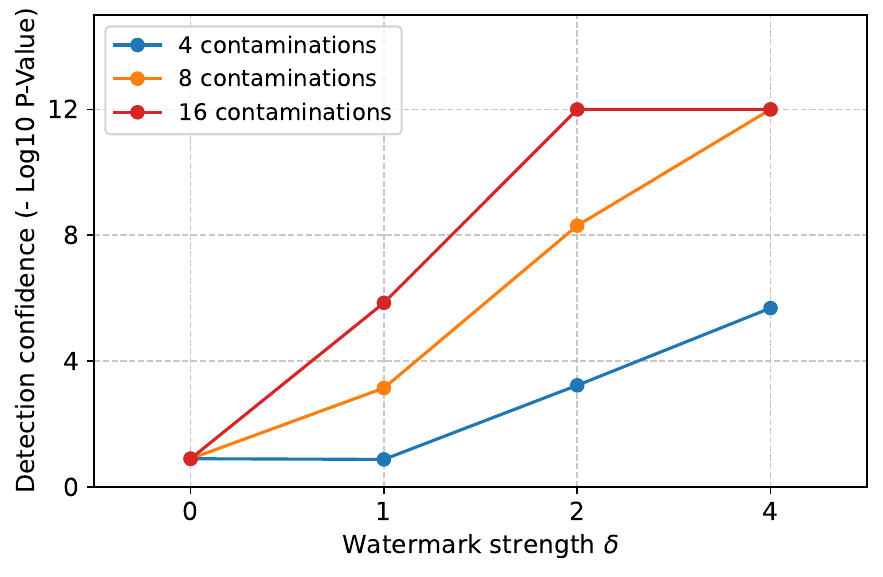}
        \subcaption{MMLU$^*$, Window size 2}
    \end{minipage}
    \vspace{-0.2cm}
    \caption{Comparison of radioactivity detection on various versions of ARC-Easy, ARC-Challenge, and MMLU$^*$ for different watermark window sizes.
    Each row corresponds to a different dataset, and each column corresponds to a different window size.
    Bigger benchmarks leads to easier detection, and window size impacts the detection confidence, the larger the better, accross all benchmarks.}
    \label{fig:appendix_watermark_contamination}
\end{figure}

\paragraph{\textbf{Ablations on different benchmarks, watermark strength, watermark window sizes, and number of contaminations.}} Results for all benchmarks (ARC-Easy, ARC-Challenge, and MMLU$^*$), with variations in watermark window size, number of contaminations, and watermark strength, are shown in \autoref{fig:appendix_watermark_performance} for utility and \autoref{fig:appendix_watermark_contamination} for radioactivity detection.
For utility, all models perform very similarly on all the rephrased versions of the benchmarks, even when pushing the watermark to $80\%$ of green tokens, although for MMLU$^*$, we observe some discrepancies. 
For instance, when using a watermarking window size of 2 (subfig i), the performance of Llama-3.2-1B increases from 38$\%$ to $42\%$ between the original and the other versions. 
However we observe the same issue when rephrasing without watermarking in that case.
The watermark window size does not have an impact.
For radioactivity detection on the other hand, as detailed in~\autoref{subsec:additional_results}, smaller window sizes correlates with lower detection confidence.


\begin{table}[t!]

    \centering

    \vspace{-0.2cm}

    \caption{
        Detection and performance metrics across different levels of contamination for ARC-Easy, ARC-Challenge, and MMLU benchmarks, watermarked with $\delta=4$.
        The performance increase is for in distribution evaluation as detailed in~\autoref{subsec:result_detection}.
        Similar results for a different templates are shown in \autoref{tab:contamination}.
    }\label{tab:contamination_indist}

    \resizebox{\textwidth}{!}{
    \begin{tabular}{r rr@{\hspace{0.5em}}l rr@{\hspace{0.5em}}l rr@{\hspace{0.5em}}l}

        \toprule

        & \multicolumn{3}{c}{ARC-Easy (1172 quest.)} & \multicolumn{3}{c}{ARC-Challenge (2373 quest.)} & \multicolumn{3}{c}{MMLU$^*$ (5000 quest.)} \\

        \cmidrule(lr){2-4} \cmidrule(lr){5-7} \cmidrule(lr){8-10}

        Contaminations & $\logpval$ & Acc. & \graydelta{\% $\Delta$} & $\logpval$ & Acc. & \graydelta{\% $\Delta$} & $\logpval$ & Acc. & \graydelta{\% $\Delta$} \\

        \midrule

        0  & -0.3 & 51.7 & \graydelta{+0.0} & -0.3 & 28.5 & \graydelta{+0.0} & -0.9 & 30.4 & \graydelta{+0.0} \\

        4  & -3.0 & 61.3 & \graydelta{+9.9} & -1.2 & 35.1 & \graydelta{+7.0} & -5.7 & 36.9 & \graydelta{+6.5} \\

        8  & -5.5 & 68.2 & \graydelta{+16.9} & -4.5 & 42.2 & \graydelta{+14.1} & \textless{-12} & 43.0 & \graydelta{+12.6} \\

        16 & \textless{-12} & 84.1 & \graydelta{+32.8} & \textless{-12} & 65.3 & \graydelta{+37.2} & \textless{-12} & 62.1 & \graydelta{+31.7} \\

        \bottomrule

    \end{tabular}
    }

    \vspace{-0.3cm}

\end{table}

\newpage

\section{Proof of correctness for contamination detection}\label{app:test-correctness-proof}

We give the proof of Proposition~\ref{proposition:fpr_p_val}. 
We remind that \(\mathcal{H}_0\) is ``The cummulative score \(S\) follows a binomial distribution \(B(\tilde{N}, 0.5)\)'' and $\text{$p$-value}(s) = \Prob(S(X_N) \geq s \mid \mathcal{H}_0) = I_{\gamma}(s+1, N-s)$ and:

\setcounter{proposition}{0}
\begin{proposition}
If we define ``being contaminated'' as having memorized the watermark, then ``not being contaminated'' matches \(\mathcal{H}_0 := S \sim B(\tilde{N}, 1/2)\).
Therefore, the test \(T_{\alpha}\) (that rejects \(\mathcal{H}_0\) if the $p$-value is less than \(\alpha\)) correctly tests for contamination, and has a False Positive Rate equal to \(\alpha\).
\end{proposition}

\begin{proof}
Assume that ``$M$ has not memorized watermark bias with secret key $\sk$''. 
Since the summed scores are i.i.d. due to de-duplication, and independent of the watermarking process because the suspect model has no other knowledge about $\sk$, and because we exclude the possibility of simply repeating watermarked tuples from the prompt through de-duplication, there is no bias towards the green or red tokens specific to $\sk$. 
Therefore, the indicators $\mathds{1} \left( y^{(t)} \text{ is in the greenlist of } \left( \sk, (x^{(t-i+1)})_{i=k}^1 \right) \right)$ are i.i.d. simulations distributed according to a Bernoulli distribution with parameter $0.5$ (in expectation over the keys). 
Thus, $S$ follows a binomial distribution $B(\tilde{N}, 0.5)$. 
So, $\mathcal{H}_0$ is true.


Reciprocally, if $\mathcal{H}_0$ is True, then there is no bias towards the green tokens, which by definition means that it has not memorized the watermark.
The $p$-value is exactly the probability to observe a score as extreme as $s$ under $H_0$, so it is the probability to observe a score as extreme as $s$ if $M$ has not memorized the watermark with secret key $\sk$ present in the benchmark.
Now let \(T_{\alpha}\) be the test that rejects \(\mathcal{H}_0\) if the $p$-value is less than \(\alpha\).
It correctly tests for contamination, and has a FPR of \(\alpha\).
\end{proof}


\vspace{-0.4cm}
\section{Compute Resources}\label{app:compute_resources}

We use our internal cluster with A-100 GPUs with 80GB memory, and:

\begin{itemize}[leftmargin=*]
    \item Each radioactivity detection test took less than 30 minutes on a single GPU. We processed the benchmark through the model, which contains a maximum of 325k tokens for MMLU$^*$ (see~\ref{tab:contamination}).
    \item Pretraining of the 1B models was conducted on 8 nodes (so 64 GPUs) and took approximately six hours. 
    Training of smaller models, with 360M and 135M parameters, was performed on 4 nodes, taking 2 hours and 1 hours respectively.
\end{itemize}

Overall, we estimate that training the 1B models required approximately 5,000 GPU hours, calculated as 3 (different window sizes) x 4 (different degrees of contamination) x 6 x 8 x 8 (GPU hours for training).
We approximate an additional factor of 2 for the other models trained, resulting in a total of approximately 10,000 GPU hours.




\clearpage

\end{document}